\documentclass{article}
\pdfoutput=1

    \usepackage[preprint, nonatbib]{neurips_2023}

\usepackage[utf8]{inputenc} 
\usepackage[T1]{fontenc}    
\usepackage{hyperref}       
\usepackage{url}            
\usepackage{booktabs}       
\usepackage{amsfonts}       
\usepackage{nicefrac}       
\usepackage{microtype}      
\usepackage{graphicx}
\usepackage{wrapfig}
\usepackage{multirow}
\usepackage[table]{xcolor}
\definecolor{gray}{rgb}{0.5,0.5,0.5}
\usepackage{color,soul}
\usepackage[toc,page,header]{appendix}

\usepackage{listings}

\usepackage{caption}
\usepackage{subcaption}

\usepackage{amsmath,amsfonts,bm}

\def\eqref#1{equation~\ref{#1}}

\def\1{\bm{1}}

\def\vx{{\bm{x}}}

\DeclareMathAlphabet{\mathsfit}{\encodingdefault}{\sfdefault}{m}{sl}
\SetMathAlphabet{\mathsfit}{bold}{\encodingdefault}{\sfdefault}{bx}{n}

\title{Deep Generative Models of Music Expectation}

\author{%
  Ninon Liz\'e Masclef \\
  \texttt{contact@ninonlizemasclef.com} \\
  \And
  T. Anderson Keller \\
  \texttt{t.anderson.keller@gmail.com}\\
}

\begin{document}

\maketitle

\begin{abstract}
A prominent theory of affective response to music revolves around the concepts of surprisal and expectation. In prior work, this idea has been operationalized in the form of probabilistic models of music which allow for precise computation of song (or note-by-note) probabilities, conditioned on a ‘training set’ of prior musical or cultural experiences. To date, however, these models have been limited to compute exact probabilities through hand-crafted features or restricted to linear models which are likely not sufficient to represent the complex conditional distributions present in music. In this work, we propose to use modern deep probabilistic generative models in the form of a Diffusion Model to compute an approximate likelihood of a musical input sequence. Unlike prior work, such a generative model parameterized by deep neural networks is able to learn complex non-linear features directly from a training set itself. In doing so, we expect to find that such models are able to more accurately represent the `surprisal' of music for human listeners. From the literature, it is known that there is an inverted U-shaped relationship between surprisal and the amount human subjects `like' a given song. In this work we show that pre-trained diffusion models indeed yield musical surprisal values which exhibit a negative quadratic relationship with measured subject `liking' ratings, and that the quality of this relationship is competitive with state of the art methods such as IDyOM. We therefore present this model a preliminary step in developing modern deep generative models of music expectation and subjective likability.
\end{abstract}

\section{Introduction}

The fields of psychology and musicology have identified expectation as a crucial factor that conveys meaning in music \cite{meyer_emotion_1956}, especially affective response \cite{juslin_everyday_2013, pearce_statistical_2018, vuust_music_2022}. The malleability of musical experience underscores that listening to music is not merely a passive activity, but an active learning process in which expectations are formed that shape our emotional responses \cite{reybrouck_neural_2021}.
Indeed, prior work has found that the sweet spot of expectation which maximizes information learning and reward system-related responses is that of intermediate complexity \cite{oudeyer_intrinsic_2016}.
Wilhelm Wundt \cite{wundt_principles_1874} proposed an inverted U-curve to describe the relationship between stimulus intensity and pleasant feeling. 
However, it was operationalized decades later and applied specifically to aesthetic pleasure by Berlyne \cite{chmiel_back_2017}. Since then, the \textit{Wundt effect} has dominated psychological research on music preference for more than two decades \cite{hargreaves_hargreaves_2010}. 
Specifically, the author \cite{berlyne_conflict_1960} found that arousal is a predominant factor in aesthetic preference, with three types of variables determining the level of arousal of a stimulus input. One in particular, collative variables, encompassing novelty, complexity, and uncertainty, has been shown to contribute most to musical liking \cite{berlyne_aesthetics_1971}. Among these collative variables, predictability has been empirically shown to contribute to music preference \cite{witek_syncopation_2014}, although this finding is not unanimous \cite{orr_relationship_2005}.
Furthermore, there is a growing body of research on serendipity as a new evaluation metric for music recommendation systems \cite{ziarani_serendipity_2021, schedl_model_2012, fu_modeling_2023}, linking surprisal and pleasure of music.

\subsection{Models of musical expectancy}

Meyer (1956) first theorized musical anticipation by applying principles of Gestalt psychology to the auditory domain to form an "emotional syntax of music" \cite{meyer_emotion_1956}. Influenced by Meyer's work, Narmour (1990) \cite{narmour_analysis_1990} developed the Implication-Realization (I-R) model of melodic expectation, which emphasized cognitive evaluation of the music rather than musical analysis of the stimuli. Instead, Narmour explained that music listening is the product of two expectation processes, one top-down and the other bottom-up \cite{narmour_top-down_1991}.
A whole tradition of predictive models of music listening was developed later in the lineage of Meyer's work. Pearce created the Information Dynamics of Music (IDyOM) model of melodic expectations \cite{pearce_construction_2005}, but replaced the rules with statistical distribution learning. IDyOM simultaneously predicts the pitch and onset time of the next note using probabilistic prediction, and calculates the final probability of the note using the joint probability of these two predictions \cite{pearce_statistical_2018}. IDyOM was effective at predicting liking from surprise, reproducing the \textit{Wundt effect} along a U-shaped curve \cite{gold_predictability_2019}. Unlike Narmour's model, IDyOM is not built from symbolic rules of musical knowledge. However, it still relies on hand-crafted features such as pitch duration. Contrary to recent diffusion models, which have billions of parameters learned from large datasets, IDyOM has only a few trainable parameters. In contrast to Narmour, Pearce argued that only top-down processing occurs, rather than both top-down and bottom-up \cite{pearce_statistical_2018}.
Furthermore, it is conditioned on the input of stimuli encoded in MIDI format. 
Therefore, musical features based on timbral or dynamic changes are not considered by the model. Instead, IDyOM focuses on syntactic features, following Meyer's argument that those features present in symbolic music are primary to the formation of musical style \cite{pearce_statistical_2018,meyer_emotion_1956}.
While symbolic music can convey emotion through expressive velocity, some affective cues are inherent in the live recording of a piece, such as \textit{rubato} \cite{todd_towards_1989}. The D-REX model was designed to address this issue with a model of surprise working on raw audio from an ecologically valid perspective of studying the music listening experience \cite{abrams_retrieving_2022}; however, the features used to compute the surprisal values in the D-REX model are restricted to learned linear transformations of the input, and therefore limited in their modeling complexity.  

Nowadays, predictive coding theories are becoming dominant in modeling human expectations over time \cite{parr_active_2022}.
The basic premise of predictive coding is that the brain operates as an expectation machine \cite{helmholtz_handbuch_1867}, constantly generating and updating predictions.
Namely, humans embody a Bayesian generative model that maps beliefs to observations. 
As a consequence, modern generative models, such as diffusion models, can be seen as instantiations of predictive coding.

Predictive coding theories claim that nervous system activity represents a process of matching internally generated predictions with external stimuli \cite{heekeren_neural_2008}. This has been demonstrated at a variety of spatial and temporal scales in the brain including music processing \cite{vuust_music_2022}, which has been crystallized in the predictive coding of music (PCM) model \cite{vuust_rhythmic_2014}. 

\section{Methods}
In this work, our goal is to leverage modern deep probabilistic generative models as measures of expectancy for music. To evaluate the quality of such a model of expectancy, we will measure the correlation structure between the output of our model --- the estimated likelihood of the data interpreted as a `surprisal' value --- and a dataset of subjective human ratings of `liking' for a set of short musical pieces. Given prior work relating music expectancy and preference, if our model is a valid model of human music expectancy, we would therefore expect to find an inverted U-shaped curve relating the surprisal and ratings. 

To approach this goal, we begin with the baseline denoising diffusion probabilistic models of Ho et al. (2020) \cite{ho_denoising_2020} as they form the foundation for much of today's state of the art in generative modeling. As we will detail in this section, due to the specific construction of these models, they admit an easily computable bound on the likelihood of the data which we will use as our approximate estimated `surprisal'. Furthermore, there exist open source implementations and model weights trained on raw musical audio, allowing for direct evaluation of such models without retraining \cite{smith_audio-diffusion_2023}. In the following section we give an overview of how such models operate, how to compute the approximate likelihood, as well as how we perform analysis to accurately compare this measure of surprisal with human listener ratings of liking.

\paragraph{Denoising Diffusion Probabilistic Models.} In brief, diffusion models can be understood as latent variable generative models composed of both a forwards and reverse process. The forward process is defined to start from the datapoint $\vx_0$ and produce sequentially noisier versions $\vx_t$ which we call the latent variables. At the end of the forward process, the distribution of final latent variables $\vx_T$ is intended to match that of the prior, typically a standard normal distribution. The reverse process then inverts this procedure, mapping from the simple prior distribution $p_{\vx_T}(\vx) = \mathcal{N}(\vx; \mathbf{0}, \mathbf{I})$ to the complex data distribution $p_{\vx_0}(\vx)$ by slowly removing noise step by step -- `denoising'. In practice, the forwards process is Markov, with the joint distribution (factorizing into a product of conditionals) and the conditional distribution of each step defined as:
\begin{equation}
    q(\{\vx_{t}\}_{t=1}^T | \vx_{0}) = \prod_{t=1}^T q(\vx_{t} | \vx_{t-1}), \hspace{10mm} q(\vx_{t} | \vx_{t-1}) = \mathcal{N}(\vx_{t}; \sqrt{\alpha_t} \vx_{t-1}, (1-\alpha_t) \mathbf{I})
\end{equation}
where $\alpha_1, \ldots, \alpha_T$ denotes the noise schedule. The reverse process is similarly Markov, and the conditional distribution which aims to remove the noise is often parameterized by a neural network with parameters $\theta$:
\begin{equation}
    p_{\theta}(\vx_{t-1} | \vx_t) = \mathcal{N}(\vx_{t-1} ; \mu_{\theta}(\vx_t, t), \Sigma_{\theta}(\vx_t, t))
\end{equation}
In practice, Ho et al. (2020) \cite{ho_denoising_2020} find that a reparamterization where the network is trained to directly predict the noise component of $\vx_t$, ($\boldsymbol{\hat{\epsilon}_{\theta}}(\vx_t, t) \approx \boldsymbol{\epsilon}$) is easier to optimize than predicting $\vx_t$ itself. 


\paragraph{Evaluating the Likelihood of Diffusion Models.} Training the parameters of denoising diffusion models can then be done by optimizing a variational bound on the negative log-likelihood:
\begin{equation}
\mathbb{E}[-\log p(\vx)] \leq \mathbb{E}_q \left[\mathcal{L}_T + \sum_{t=1}^T \mathcal{L}_{t-1} - \mathcal{L}_0 \right]
\end{equation}
where $\mathcal{L}_0 = \log p(\vx_0 | \vx_1)$ is the reconstruction loss of the true data given $\vx_1$; $\mathcal{L}_T = D_{KL}\left(q(\vx_T | \vx_0) || p_{\vx_T}(\vx)\right)$ is the prior loss, measuring the distance of the final latent variable from the prior; and $\mathcal{L}_t$ is the diffusion loss over all time steps defined below. Following this, Kingma et al. (2023) \cite{kingma_variational_2023} show that the diffusion loss term can be expressed simply as:
\begin{equation}
   \mathcal{L}_t(\vx) =  \mathbb{E}_{\boldsymbol{\epsilon} \sim \mathcal{N}(\mathbf{0}, \mathbf{I})} \left[\frac{1}{2} \left(1 - \mathrm{SNR}(t-1) / \mathrm{SNR}(t) \right) || \boldsymbol{\epsilon} - \boldsymbol{\hat{\epsilon}_{\theta}}(\vx_t, t)||^2\right]
\end{equation}
where $\mathrm{SNR}(t) = \frac{\alpha^2_t}{1- \alpha^2_t}$. Further, following \cite{hoogeboom_equivariant_2022} we know that $\mathcal{L}_{T}$ is close to zero when the noise schedule is defined such that $\alpha_T \approx 0$. In this work, we take this bound on the likelihood as our approximate measure of expectancy or surprisal for each music sample in our dataset. 

\paragraph{Data.} To evaluate the diffiusion models, we make use of the dataset from Gold et al. \cite{gold_predictability_2019} which contains 57 audio files, liking ratings from 44 subjects, and the associated output of the IDyOM model (mean duration weighted information content, mDW-IC) which we use as a baseline.
Since we are working with pre-trained diffusion models which operate on 5-second audio clips, we simply break each clip into 5-second non-overlapping blocks and take the sum of the estimated likelihoods as the total likelihood value. Although this is non-ideal and introduces significant boundary effects, we find in practive it works sufficiently well for this study.


\paragraph{Analysis.} To compare our surprisal value with liking ratings, we follow the procedure of Gold et al. \cite{gold_predictability_2019}, first fitting a linear mixed effects model to the data (python $\mathrm{statsmodels}$'s $\mathrm{mixedlm}$) and using this to adjust each subjects rating based on their estimated individual random effects. This allows to us account for the subjective properties of likability rating per user, and compare them to our model which only produces a single likelihood value per audio clip. We then compute the goodness of fit of a quadratic function mapping from surprisal values to adjusted rating using ordinary least squares.
\section{Experiments}
In this section we report the results of evaluating the pretrained Audio-Diffusion model (\textit{audio-diffusion-256}) from Smith \cite{smith_audio-diffusion_2023} on the Gold et al. (2019) \cite{gold_predictability_2019} audio dataset. The model is trained on a dataset of over 20,000 mel spectrograms of resolution 256x256 computed using a sampling rate of 22050 Hz, a window length of 2048, and a stride of 512.

\begin{figure}[h!]
\vspace{-2mm}
    \centering
\includegraphics[width=0.95\textwidth]{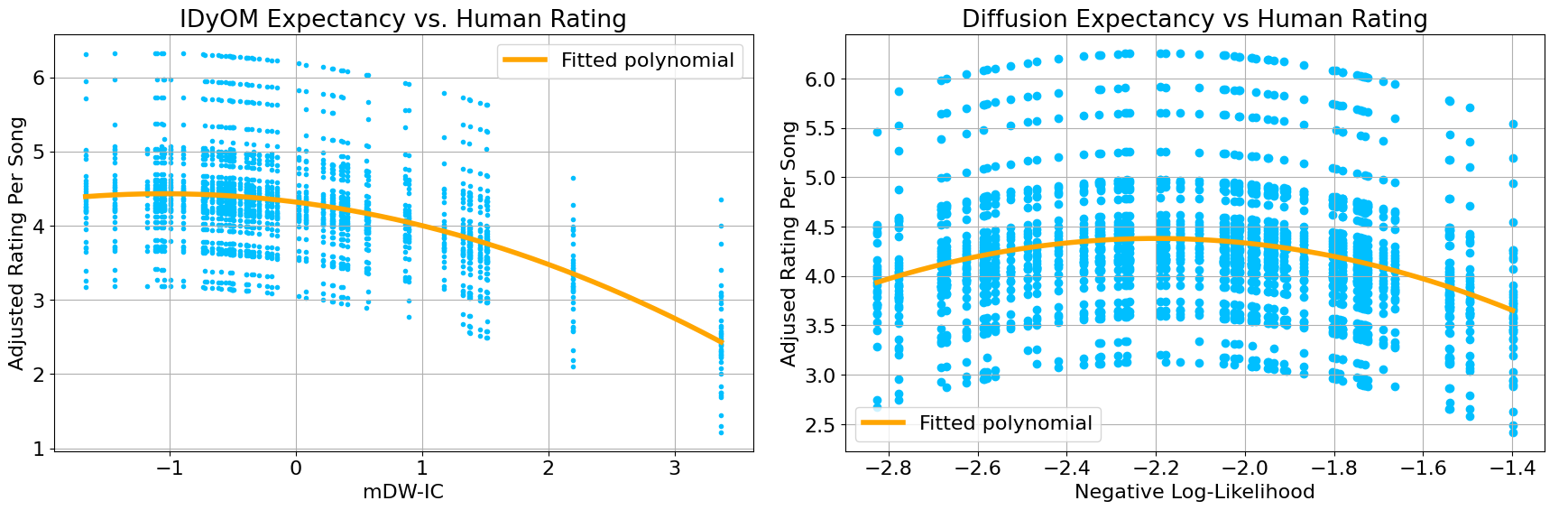}
    \caption{Plot of the observed Wundt effect for the IDyOM model (left) and our proposed diffusion model (right). Data points for the adjusted human-subject ratings of each song are plotted in blue, with the best quadratic fit model plotted in orange.}
    \label{fig:wundt}
    \vspace{-4mm}
\end{figure}

\paragraph{Measuring the \textit{Wundt Effect}.} As a primary result of this study, we seek to discover if an inverted U-shape relationship exists between the surprisal values computed by a diffusion model and the human liking ratings from the dataset. In Figure \ref{fig:wundt}, we plot the results of our analysis for both the baseline IDyOM model (left) and the Audio Diffusion model (right). We see that indeed the Wundt effect is present, with the diffusion model exhibiting a further rightward shift of peak likeability compared with IDyOM. From the literature \cite{gold_predictability_2019}
, this shift is consistent with expertise and perhaps is a result of the diffusion model's more diverse training set compared with  the IDyOM counterpart.


\paragraph{Goodness of Fit.} To evaluate the quality of the above fit, in Table \ref{tab:good} we evaluate standard goodness of fit measures compared with the IDyOM model (such as $R^2$, Log-likelihood, AIC and BIC). We see that fitted polynomial resulting from the diffusion model output achieves a higher likelihood, and better AIC/BIC values, with a lower $R^2$ value due the the significant inter-subject variability per song. Despite this lower $R^2$, we see that the coefficient of the quadratic term of the polynomial is indeed negative and the fit is highly significant (coeff=$-1.135$, 95\% CI [$-1.573$ $-0.697$], $p<1e-4$). 
\begin{table}[h!]
\vspace{-3mm}
  \caption{Goodness of fit measures for the IDyOM model and the diffusion model. We see that while the IDyOM model has a higher $R^2$, indicating greater variance of the data is explained, the Log-likelihood of the diffusion model is markedly lower.}
  \label{tab:good}
  \vspace{3mm}
  \centering
  \begin{tabular}{l|cccc}
    \toprule
    Surprisal Model     & $R^2$  & Log-Likelihood & AIC & BIC \\
    \midrule
    IDyOM & \textbf{0.240} &  -1869.0  & 3744 & 3761 \\
    Audio-Diffusion &  0.062   & \textbf{-1860.5} & \textbf{3727} & \textbf{3744}  \\
    \bottomrule
  \end{tabular}
  \vspace{-5mm}
\end{table}


\section{Discussion}

This work is proposed primarily as a proof of concept that modern deep generative models can be used to model music expectancy.
We believe that the experiments presented in this study, while not fully exploiting the power of state of the art diffusion models, adequately demonstrate that surprise scores from diffusion models do exhibit the \textit{Wundt effect}, and thus validate the idea that diffusion models can be used to model music expectancy.
The potential advantages of diffusion models over existing expectation models, e.g. IDyOM and D-REX, on a practical level are the larger numbers of trainable parameters, allowing the ingestion of larger datasets, as well as the use of deep feature extractors which should be able to more accurately model complex distributions.
Future research endeavors should involve replicating the experiment training on a variety of datasets, especially non-Western music, and studying the interaction of unexpectedness with other collative variables such as complexity and familiarity. Furthermore, as demonstrated by the low $R^2$ values of our experiments, intersubject variability is significantly challenging to capture with a single model. Significant future work should be directed towards training of individual subject-level models of expectancy, either through fine-tuning or self-supervised training. 
As a result, an application of this work could be the construction of an individual profile of serendipitous music recommendation systems \cite{schedl_model_2012}. In the same way as diffusion models were used by Wang \cite{wang_diffusion_2023} for music recommendation, our measure of musical expectation could predict the user's liking profile from the audio level instead of the user's listening trajectory.
In addition, it would be useful to compare the likelihood score of the diffusion model with biosignal correlates of surprise, such as event-related potentials (ERPs), as done in \cite{abrams_retrieving_2022}.


\bibliographystyle{plain}
\bibliography{neurips_2023}


\end{document}